\documentclass{article}

\usepackage[preprint]{neurips_2026}

\usepackage[utf8]{inputenc} % allow utf-8 input
\usepackage[T1]{fontenc}    % use 8-bit T1 fonts
\usepackage{hyperref}       % hyperlinks
\usepackage{url}            % simple URL typesetting
\usepackage{booktabs}       % professional-quality tables
\usepackage{amsfonts}       % blackboard math symbols
\usepackage{nicefrac}       % compact symbols for 1/2, etc.
\usepackage{microtype}      % microtypography
\usepackage{xcolor}         % colors
\usepackage{graphicx}
\usepackage{caption}
\usepackage{fontawesome5}
\workshoptitle{Representations for the Physical Sciences Workshop}
\title{DINOspec: Efficient Multimodal Alignment of Vision and Spectral Foundation Models for Astronomy}

\author{%
  Erica~Lastufka \\
    University of Geneva\\
  Geneva, Switzerland 1211 \\
  \texttt{erica.lastufka@unige.ch} \\
  \And
  Mariia Drozdova \\
  University of Geneva \\
  Geneva, Switzerland 1211 \\
  \texttt{mariia.drozdova@unige.ch} \\
    \And
  Daniel Schaerer \\
  University of Geneva \\
  Geneva, Switzerland 1211 \\
  \texttt{daniel.schaerer@unige.ch} \\
  \AND
  Svyatoslav Voloshynovskiy \\
  University of Geneva \\
  Geneva, Switzerland 1211 \\
  \texttt{Svyatoslav.Voloshynovskyy@unige.ch} \\
}

\begin{document}

\maketitle

\begin{abstract}
Astronomical observations provide multimodal views of physical systems, with images and spectra capturing complementary properties of celestial objects. Scientific foundation models can learn powerful representations from these observations, but representations learned by separate models remain difficult to combine. We investigate whether laten space representations learned by separate vision and spectral models can be aligned without retraining their encoders. We introduce DINOspec, a multimodal framework that aligns a frozen DINOv3 image encoder with a pre-trained AION-1 spectral encoder using lightweight adapters and contrastive learning on 20,472 paired images and spectra of astronomical objects. DINOspec improves galaxy morphology classification (F1: 0.72$\rightarrow$0.78) and spectral type classification (F1: 0.70$\rightarrow$0.74) while training at few as 21M parameters. 
 Redshift prediction shows no consistent improvement over the AION-1 baseline of $R^2\approx0.9$, demonstrating that potential benefits of alignment depends on the configuration and downstream task. These results provide an empirical study of composing astronomical representations from independently pretrained foundation models through lightweight alignment modules.
%In contrast, spectroscopic redshift prediction shows no consistent improvement ($R^2\approx0.9$), demonstrating that. These results demonstrate that scientific foundation models can be composed through lightweight representation alignment.

\end{abstract}

\section{Introduction}

Observational astrophysics provides a natural setting for studying multimodal representation learning, as different instruments measure complementary aspects of the same physical systems. Multi-band imaging, spectroscopy, photometry, and source catalogs each contain different information about astronomical objects, but combining these observations requires representations compatible across modalities. Recent astronomical foundation models have begun addressing this challenge. AstroCLIP \citep{parker_astroclip_2024} aligns optical images and spectra, AstroM3 \citep{rizhko_astrom3_2025} extends contrastive learning to photometry, spectra, and metadata, and AION-1 \citep{parker_aion-1_2025} integrates five heterogeneous datasets using modality-specific tokenizers and cross-modal transformer layers.

A central challenge for scientific foundation models is that representations learned from different instruments or datasets are difficult to reuse and combine. Astronomical observations differ substantially in wavelength coverage, spatial resolution, and noise properties, producing domain shifts between instruments or even surveys. Existing multimodal architectures typically rely on jointly trained modality-specific components, particularly tokenizers optimized for the statistics of the training data. Even unimodal benchmarks show that visual encoders of various architectures perform differently across data regimes \citep{lastufka_examining_2025}. Consequently, incorporating new data sources into existing multimodal models would require retraining substantial portions of it rather than reusing existing physical representations.

In this work, we investigate whether independently learned scientific representations can be aligned through lightweight trainable modules. This separates representation learning from cross-modal alignment, allowing use of existing foundation models without modifying their learned representations. We build on the Locked-image Tuning (LiT) strategy introduced by DINO.txt \citep{jose_dinov2_2024}, which demonstrated that frozen visual representations can be adapted to new semantic spaces using  alignment layers. Our method, DINOspec, combines a frozen DINOv3 \citep{simeoni_dinov3_2025} visual encoder with a spectral encoder operating on tokens from the pre-trained AION-1 tokenizer. 
%new
DINOspec does not introduce a new contrastive alignment objective; instead, it investigates the empirical behavior of lightweight alignment between independently pretrained astronomical image and spectral representations. We evaluate how adapter initialization, capacity, and training strategy affect downstream morphology classification, spectral classification, and redshift regression, providing evidence of when alignment improves downstream utility and when unimodal information already dominates the downstream task.
%We investigate asymmetric knowledge transfer between frozen encoders by varying the initialization and capacity of the trainable adapters, studying when multimodal alignment improves scientific representations and when unimodal information already dominates the downstream task.

\section{Data}\label{sec:data}

\begin{table*}[h]
\centering
\resizebox{\textwidth}{!}{
\begin{tabular}{llllll}%{p{2.3cm}p{1.9cm}p{1.1cm}p{3.4cm}p{3.3cm}}
\toprule
Dataset name & Origin & \raggedright{Train} & Test & Labels & Task \\
\midrule
%MiraBest & 750 & sparse, centered on bright galaxy & binary classification \\
hsc & HSC \textit{grz} images & 86614 &  & &vision pre-training\\ %MMU/hsc
sdss\_desi & SDSS and DESI spectra & 107223 & 102407 &  & spectra pre-training\\
desi\_xmatchGRZ & cross-matched HSC and DESI & 16378 & 4094 &  & vision and joint pre-training\\
GZ10\_DECals & LegacySurvey \textit{grz} images & 14189 & 3546 & 10 classes & classification\\
GZ10\_xmatch & HSC \textit{grz} images & 553 & 139 & 10 classes & classification\\
%GZ10 subset & DeCALS images & 5652 & 1413 & 10 classes & classification \\
SDSS16\_reg & SDSS spectra & 14000 & 3000 & redshift & regression \\
SDSS16\_class & SDSS spectra & 182000 & 39000 & spectral class & classification \\
\bottomrule
\end{tabular}
}
\caption{\label{tab:datasets}The datasets used in this study.}
\end{table*}

For cross-modal alignment, we use paired imaging and spectroscopic observations of the same astrophysical objects. The pre-trained image and spectral encoders are learned from larger unimodal datasets (LVD-1689M for DINOv3 and SDSS/DESI data for AION-1), while we additionally investigate whether unimodal pre-training the alignment blocks improves joint representations.
The datasets used for unimodal pre-training (when applicable), joint multi-modal alignment, and downstream task evaluation (described in Section \ref{sec:evaluation}) are summarized in Table~\ref{tab:datasets}. Further details of dataset construction are in Appendix \ref{sec:app_data}.

Image data originate from the Hyper Suprime-Cam's (HSC) deep and ultra-deep surveys \citep{aihara_hyper_2018}, which provides calibrated optical imaging in five bands ($g$, $r$, $i$, $z$, and $y$). We use the $g$, $r$, and $z$ bands which cover a broad redshift range to construct RGB images using the \citet{lupton_preparing_2004} \textit{asinh} intensity mapping. Image cutouts are extracted at a pixel scale of 0.162 arcsec pixel$^{-1}$, centered on cataloged sources, with a final size of $160\times160$ pixels. A total of 86,614 such cutouts were obtained.

Optical spectra were from the Sloan Digital Sky Survey (SDSS; \citet{ahumada_16th_2020}) and the Dark Energy Spectroscopic Instrument (DESI; \citet{collaboration_desi_2016}). SDSS provides medium-resolution optical spectra covering approximately 3650--10400 \AA, while DESI provides spectra spanning approximately 3600--9800 \AA{} with wavelength-dependent resolution. The combined SDSS and DESI dataset contained 209630 object spectra.

Finally, we constructed cross-matched HSC and DESI objects within a one arcsecond radius to obtain a dataset containing 20,472 cutouts. These paired observations are used for CLIP-style contrastive training, to align image and spectrum data in a shared representation space. 
%new
Objects from this cross-matched dataset were excluded from any downstream task evaluation datasets in order to prevent leakage.

\section{Models}

\begin{figure*}[t]
\begin{minipage}[t]{0.46\textwidth}
\vspace{0pt}
%\centering
\includegraphics[width=\linewidth]{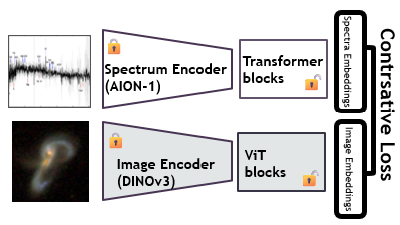}
\label{fig:diagram}
\caption{The DINOspec architecture.}
\end{minipage}
\hfill
\begin{minipage}[t]{0.46\textwidth}
\vspace{0pt}
%\centering
\begin{tabular}{lc}
\toprule
Model Component & Parameter Count \\
\midrule
%DINOv3-B backbone & 86M \\
DINOv3 Large \small\faLock & 300M \\
AION-1 Base \small\faLock & 300M \\\midrule
ViT block & 7M \\
2x ViT block & 14M \\
Embedding layer & 1.18M \\
2 block Transformer & 6.9M \\
8 block Transformer & 27.8M \\
16 block Transformer & 55.6M \\
2-layer MLP & 0.33M \\
\bottomrule
\end{tabular}
\captionof{table}{\label{tab:params}Model parameter counts.}
\end{minipage}
\end{figure*}

DINOspec adopts the structure introduced in DINO.txt \citep{jose_dinov2_2024}, consisting of a frozen DINOv3-Large vision backbone and a spectral branch operating on tokens from the frozen AION-1 tokenizer. The vision branch appends transformer blocks to the DINOv3 output, while the spectral branch uses either a linear embedding layer or additional transformer blocks. The two branches are connected through trainable adapters, which align visual and spectral representations using the symmetric cross-entropy CLIP loss \citep{radford_learning_2021}.
Table \ref{tab:params} shows different components of the DINOspec model which we investigated in this work, and the number of trainable parameters in each component. For comparison, the parameter counts of the DINOv3 and AION-1 models used are also included, although these models are used exclusively as frozen encoders.

We designed seven experimental configurations (Table~\ref{tab:configs}) to investigate multimodal alignment under different assumptions about the relative strength of the vision  and spectral encoders. Here, pre-training refers to unimodal training of randomly initialized transformer blocks using either the BERT reconstruction objective or the DINOv3 loss; details are provided in the Appendix.

Configuration 0 uses pre-trained vision blocks with a lightweight spectral embedding layer, while Configuration 1 introduces a pre-trained spectral model and evaluates alignment when the spectral branch guides the visual representation. Configurations 2 and 3 reverse this assumption, using pre-trained visual representations while varying the initialization of the spectral adapter. Finally, Configurations 4--6 make no specific assumption about the strength of either encoder, and compare different combinations of pre-trained and randomly initialized components when all trainable modules are optimized jointly. For each configuration, joint training was performed using the cross-matched dataset for 150 epochs and a learning rate of $5e-6$.

\begin{table}[t]
\centering
\resizebox{\textwidth}{!}{
\begin{tabular}{clllp{4.5cm}}
\toprule
Configuration \# & Assumption & Vision blocks & Transformer blocks & Training strategy \\
\midrule
0 & Strong spectra 
  & Pre-trained 
  & Embedding layer only 
  & Joint training \\

1 & Strong spectra 
  & Random 
  & Pre-trained  
  & SM frozen for first 50 epochs \\

2 & Strong vision 
  & Pre-trained 
  & Pre-trained 
  & VM frozen for first 50 epochs \\

3 & Strong vision 
  & Pre-trained 
  & Random 
  & VM frozen for first 50 epochs \\

4 &  
  & Pre-trained 
  & Random 
  & Joint training \\

5 &  
  & Pre-trained 
  & Pre-trained  
  & Joint training \\

6 &  
  & Random 
  & Pre-trained 
  & Joint training \\

\bottomrule
\end{tabular}
}
\caption{Experiment configurations. The vision backbone and spectral tokenizer are always frozen. The vision backbone plus any additional blocks compose the entire vision model (VM); likewise for the spectral model (SM). Initialization of additional blocks is either “Pre-trained” or “Random”.}
\label{tab:configs}
\end{table}

\section{Evaluation}\label{sec:evaluation}

We evaluate the utility of aligned representations on both vision and spectral downstream tasks rather than direct measures of alignment such as retrieval or cross-modal matching. 
%We evaluate the utility of aligned representations through downstream vision and spectral tasks rather than using retrieval or cross-modal matching as the primary evaluation. 
Images and spectra provide complementary, but not identical, information about astronomical objects; retrieval performance alone therefore does not establish whether an alignment preserves the modality-specific information required for scientific inference. Our evaluation measures downstream utility, rather than simply alignment quality.

Galaxy morphology classification used the 10-class Galaxy10 DECaLS dataset \citep{leung_henryskygalaxy10_2026}. Since AION-1 requires a specific image preprocessing pipeline, we cross-matched Galaxy10 objects with available HSC imaging, resulting in 692 usable images (GZ10\_xmatch in Table~\ref{tab:datasets}). For consistency, AION-1 was evaluated using only HSC images and no additional data modalities. Unless otherwise specified, image morphology classification  used only the vision encoder and a 2-layer MLP classifier.

%new
As a complementary assessment on a larger dataset and imaging source than the HSC subset used for the primary morphology evaluation, we evaluate morphology classification on the full Galaxy10 DECaLS dataset using Legacy Survey images. LegacySurvey images are have a larger pixel scale than HSC images, resulting in lower resolution, and the survey prioritized coverage over depth. Together with different response profiles in the same imaging bands, these galactic images have slightly different properties than the joint training distribution DINOspec models were exposed to. This experiment should be treated as a qualitative check of whether the results observed in the primary evaluation persist on a larger dataset, rather than as a controlled comparison of performance across imaging domains. Full results of this evaluation are in Appendix \ref{sec:full_results}.
%In this experiment, we compare the DINOv3-based model with trained alignment blocks against the AION-1 Legacy Survey image encoder baseline. Because the datasets differ in both imaging domain and sample composition, this comparison is not a controlled test of domain shift alone.

Spectral tasks used the SDSS16 dataset \citep{zhong_galaxy_2024}, which provides labels for redshift and spectral class. We evaluated redshift regression for galaxies and classification of stars, galaxies, and quasars (SDSS16\_reg and SDSS16\_class in Table~\ref{tab:datasets}). As for vision tasks, evaluation used only the corresponding spectral encoder with a 2-layer MLP classifier, modified for classification or regression.

Learning rates were selected through a small hyperparameter sweep for each task, with early stopping applied. Each evaluation was repeated seven times with different random seeds. We report F1 scores for classification and the coefficient of determination ($R^2$) for redshift prediction.

\begin{table*}[h]
\resizebox{\textwidth}{!}{
\begin{tabular}{rllllll}
\toprule
\# & Transformer  & Pre-trained  & Trainable  & Image F1 & Spectra F1 & Spectra R$^2$ \\
& Blocks &  ViT Blocks? &  Parameters &  &  &  \\
\midrule
1 & 2* & N + LoRA & 21.4M &$\mathbf{0.796 \pm 5.8e-3}$ & $0.708 \pm 3.4e-3$ & $0.879 \pm 1.7e-2$ \\
1 & 2* & N & 21M & $0.779 \pm 6.1e-3$ & $0.704 \pm 5.7e-3$ & $0.882 \pm 1.4e-2$ \\
1 & 16* & N & 70M & $0.773 \pm 5.5e-3$ & $0.651 \pm 1.4e-3$ & $0.893 \pm 1.0e-3$ \\
6 & 16 & N & 70M & $0.768 \pm 5.5e-3$ & $0.659 \pm 3.5e-3$ & $0.891 \pm 5.5e-4$ \\
6 & 8 & N & 42M & $0.764 \pm 5.3e-3$ & $0.663 \pm 4.7e-3$ & $0.895 \pm 1.0e-3$ \\
1 & 8* & N & 42M & $0.758 \pm 5.3e-3$ & $0.667 \pm 5.5e-3$ & $0.895 \pm 6.2e-4$ \\
%0 & - & N & $0.752 \pm 5.7e-3$ & $0.697 \pm 1.0e-3 $ & $0.910 \pm 4.8e-3$ \\
6 & 2 & N & 21M & $0.747 \pm 4.6e-3$ & $0.688 \pm 3.9e-3$ & $0.884 \pm 1.6e-2$ \\
2 & 2 & Y* & 21M & $0.732 \pm 4.2e-3$ & $\mathbf{0.736 \pm 2.8e-3}$ & $0.879 \pm 3.9e-3$ \\
4 & 16\^ & Y & 70M & $0.731 \pm 6.1e-3$ & $0.732 \pm 1.4e-3$ & $0.860 \pm 4.4e-3$ \\
0 & - & Y & 15M & $0.731 \pm 6.0e-3$ & $0.697 \pm 1.0e-3 $ & $0.910 \pm 4.8e-3$ \\
5 & 8 & Y & 42M & $0.728 \pm 8.6e-3$ & $0.711 \pm 4.5e-3$ & $0.898 \pm 1.1e-3$ \\
4 & 8\^ & Y & 42M & $0.728 \pm 6.2e-3$ & $0.717 \pm 2.2e-3$ & $0.891 \pm 5.9e-3$ \\
5 & 2 & Y & 21M & $0.724 \pm 5.7e-3$ & $0.706 \pm 2.1e-3$ & $0.890 \pm 3.0e-3$ \\
3 & 2\^ & Y* & 21M & $0.724 \pm 5.6e-3$ & $0.714 \pm 5.1e-3$ & $0.873 \pm 2.3e-3$ \\
% 2 & 8 & Y* & $0.722 \pm 6.0e-3$ & $0.691 \pm 2.1e-2$ & $0.899 \pm 1.2e-3$ \\
% 4 & 2\^ & Y & $0.721 \pm 4.5e-3$ & $0.709 \pm 4.1e-3$ & $0.888 \pm 2.0e-3$ \\
% 5 & 16 & Y & $0.720 \pm 5.9e-3$ & $0.653 \pm 3.7e-3$ & $0.884 \pm 1.1e-3$ \\
% 3 & 8\^ & Y* & $0.717 \pm 8.5e-3$ & $0.731 \pm 2.8e-3$ & $0.874 \pm 6.1e-3$ \\
\midrule
& \multicolumn{1}{r}{DINOv3 ViT-L} & + LoRA + MLP & 15.33M & $0.757 \pm 1.9e-2$ & - & - \\
& \multicolumn{1}{r}{DINOv3 ViT-L} & + MLP & 0.33M & $0.722 \pm 5.2e-3$ & - & - \\
& \multicolumn{1}{r}{AION-1 B} & + MLP & 0.33M & $0.702 \pm 1.2e-2$ & $0.697 \pm 1.0e-3 $ & $\mathbf{0.910 \pm 4.8e-3}$ \\
\bottomrule
\end{tabular}
}
\caption{\label{tab:results}Evaluation results from joint pre-training (* frozen model first 50 epochs, \^{}randomly initialized). Baseline comparisons to vision (DINOv3) and multimodal (AION-1) models included.}
\end{table*}

\section{Results}

Table~\ref{tab:results} summarizes performance across image and spectral tasks. DINOspec improves morphology and spectral classification over the DINOv3-L and AION-1 baselines while maintaining similar redshift prediction performance. When the spectral encoder is used to guide alignment with the image branch (Configurations 1 and 6), galaxy morphology classification sees the strongest improvements, from the baseline F1=0.72 to F1=0.78, which out-performs DINOv3 even after LoRA fine-tuning. In contrast, spectral classification improves the most from F1=0.70 to F1=0.74 when alignment is performed after pre-training of the extra vision and spectral blocks, keeping the entire image branch frozen for the first 50 epochs of joint training and allowing the visual features to guide alignment. Notably, the same pre-trained components when trained jointly without restriction (Configuration 5) do not result in the same magnitude of improvement in either morphology or spectral classification.

Visual galaxy morphology classification is the task that benefits most from multimodal joint training. As there exists some domain shift given that the DINOv3 backbone was likely not exposed to significant samples of different galaxy types during pre-training, this is perhaps to be expected.
%new
The top result, that Configuration 1 with the pre-trained spectral model performs best on morphology classification, is confirmed by our additional evaluation of the full Galaxy10 DECaLS dataset using Legacy Survey images, as is the general finding that multimodal joint training benefits downstream morphology classification over baselines.
%The DINOv3-based model with alignment blocks trained on HSC data outperforms the AION-1 Legacy Survey image encoder baseline in this setting. This indicates that the DINOv3-based representation remains effective under this cross-survey evaluation. However, because the models differ in pretraining, architecture, and downstream adaptation, and the datasets differ in sample composition, this result does not establish that DINOv3 representations are generally more flexible or isolate the effect of domain shift.
 LoRA adaptation of the vision backbone for that best-performing configuration achieved F1=0.796 on HSC images. Since this requires updating an additional $\approx 0.6M$ parameters, it is not directly comparable to the other configurations but demonstrates an efficient alternative for improving visual performance.
Increasing the number of spectral adapter parameters did not consistently improve performance. This suggests that the AION-1 spectral encoder already contains sufficient information for alignment, a hypothesis that is supported by the fact that using only an embedding layer instead of multiple transformer blocks still results in an improvement in visual morphology classification.

%Results show that, unlike morphology and spectral classification, redshift prediction remains close to the AION-1 baseline. 
Unlike morphology and spectral classification, redshift prediction shows no consistent improvement over the AION-1 baseline (\(R^2=0.910\)). Some configurations approach this performance, while others show a reduction in \(R^2\).
A possible explanation is that the spectral encoder already captures the information required for redshift estimation, while the image modality provides little additional signal. % for this task. 
The use of Lupton-scaled $grz$ images, which removes the exact flux ratios needed for photometric redshift estimation, may also limit the usefulness of images for redshift estimation. 

Together, these results show that lightweight alignment can improve downstream performance without jointly training large multimodal encoders, provided sufficiently strong unimodal representations exist. 
%The success of DINOv3 as a vision backbone further suggests that domain-specific encoders are not always required; instead, domain adaptation can be achieved through the alignment modules, although this may require additional trainable capacity.
The strong performance of DINOv3 across the HSC and Legacy Survey morphology evaluations also suggests that general-purpose vision backbones pretrained on natural images can be useful for astronomical applications, potentially offering greater flexibility across imaging surveys than more specialized models like AION-1. This highlights the potential of alignment modules to adapt broadly pretrained visual representations to astronomical data without requiring a dedicated vision encoder for each instrument or survey.
%new
 % This suggests that the alignment strategies evaluated here do not reliably improve redshift prediction, potentially because the spectral representation already captures much of the information needed for the task. The use of Lupton-scaled \(grz\) images, which do not preserve the original band flux ratios directly, may also limit the usefulness of the image modality for redshift estimation.

\section{Conclusion}

The increasing diversity of astrophysical surveys and instruments makes it impractical to rely exclusively on jointly trained multimodal architectures. Our results demonstrate that representation learning and cross-modal alignment can be separated, allowing independently developed foundation models to be reused. Adapter-based alignment using as few as 21M parameters successfully combines independently pre-trained vision and spectral models while preserving their individual capabilities. Improvements in morphology and spectral classification shows alignment is most effective when modalities contain complementary information about the physical object. Conversely, unchanged redshift prediction demonstrates that alignment does not provide additional information when one modality already contains the dominant signal for a task. The benefit of multimodal representation learning therefore 
depends on the relationship between modalities and the objectives.

DINOspec extends representation alignment methods to astronomical observations, where different instruments capture distinct physical processes across the electromagnetic spectrum. Rather than requiring a new jointly trained architecture for each combination of modalities, it provides a framework for connecting independently developed scientific foundation models through alignment modules each containing only $\mathcal{O}(10^6)$ parameters.

\newpage
\bibliographystyle{plainnat}
\bibliography{dinospec}

@inproceedings{radford_learning_2021,
	title = {Learning transferable visual models from natural language supervision},
	booktitle = {International conference on machine learning},
	publisher = {PMLR},
	author = {Radford, Alec and Kim, Jong Wook and Hallacy, Chris and Ramesh, Aditya and Goh, Gabriel and Agarwal, Sandhini and Sastry, Girish and Askell, Amanda and Mishkin, Pamela and Clark, Jack and {others}},
	year = {2021},
	pages = {8748--8763},
}

@misc{jose_dinov2_2024,
	title = {{DINOv2} {Meets} {Text}: {A} {Unified} {Framework} for {Image}- and {Pixel}-{Level} {Vision}-{Language} {Alignment}},
	shorttitle = {{DINOv2} {Meets} {Text}},
	url = {http://arxiv.org/abs/2412.16334},
	doi = {10.48550/arXiv.2412.16334},
	language = {en},
	urldate = {2025-01-14},
	publisher = {arXiv},
	author = {Jose, Cijo and Moutakanni, Théo and Kang, Dahyun and Baldassarre, Federico and Darcet, Timothée and Xu, Hu and Li, Daniel and Szafraniec, Marc and Ramamonjisoa, Michaël and Oquab, Maxime and Siméoni, Oriane and Vo, Huy V. and Labatut, Patrick and Bojanowski, Piotr},
	month = dec,
	year = {2024},
	note = {arXiv:2412.16334 [cs]},
}

@article{rizhko_astrom3_2025,
	title = {{AstroM3}: {A} {Self}-supervised {Multimodal} {Model} for {Astronomy}},
	volume = {170},
	issn = {1538-3881},
	shorttitle = {{AstroM3}},
	url = {https://dx.doi.org/10.3847/1538-3881/adcbad},
	doi = {10.3847/1538-3881/adcbad},
	language = {en},
	number = {1},
	urldate = {2025-09-09},
	journal = {AJ},
	publisher = {The American Astronomical Society},
	author = {Rizhko, M. and Bloom, J. S.},
	month = jun,
	year = {2025},
	pages = {28},
}

@article{zhong_galaxy_2024,
	title = {Galaxy {Spectra} neural {Network} ({GaSNet}). {II}. {Using} deep learning for spectral classification and redshift predictions},
	volume = {532},
	issn = {0035-8711},
	url = {https://doi.org/10.1093/mnras/stae1461},
	doi = {10.1093/mnras/stae1461},
	number = {1},
	urldate = {2026-01-14},
	journal = {Mon Not R Astron Soc},
	author = {Zhong, Fucheng and Napolitano, Nicola R and Heneka, Caroline and Li, Rui and Bauer, Franz Erik and Bouche, Nicolas and Comparat, Johan and Kim, Young-Lo and Krogager, Jens-Kristian and Longhetti, Marcella and Loveday, Jonathan and Roukema, Boudewijn F and Rouse, Benedict L and Salvato, Mara and Tortora, Crescenzo and Assef, Roberto J and Cassarà, Letizia P and Costantin, Luca and Croom, Scott M and Davies, Luke J M and Fritz, Alexander and Guiglion, Guillaume and Humphrey, Andrew and Pompei, Emanuela and Ricci, Claudio and Sifón, Cristóbal and Tempel, Elmo and Zafar, Tayyaba},
	month = jul,
	year = {2024},
	pages = {643--665},
}

@article{lupton_preparing_2004,
	title = {Preparing {Red}-{Green}-{Blue} {Images} from {CCD} {Data}},
	volume = {116},
	issn = {0004-6280},
	url = {https://ui.adsabs.harvard.edu/abs/2004PASP..116..133L},
	doi = {10.1086/382245},
	urldate = {2026-03-18},
	journal = {Publications of the Astronomical Society of the Pacific},
	publisher = {IOP},
	author = {Lupton, Robert and Blanton, Michael R. and Fekete, George and Hogg, David W. and O'Mullane, Wil and Szalay, Alex and Wherry, Nicholas},
	month = feb,
	year = {2004},
	note = {ADS Bibcode: 2004PASP..116..133L},
	pages = {133--137},
}

@article{aihara_hyper_2018,
	title = {The {Hyper} {Suprime}-{Cam} {SSP} {Survey}: {Overview} and survey design},
	volume = {70},
	issn = {0004-6264},
	shorttitle = {The {Hyper} {Suprime}-{Cam} {SSP} {Survey}},
	url = {https://doi.org/10.1093/pasj/psx066},
	doi = {10.1093/pasj/psx066},
	number = {SP1},
	urldate = {2026-07-07},
	journal = {Publ Astron Soc Jpn Nihon Tenmon Gakkai},
	author = {Aihara, Hiroaki and Arimoto, Nobuo and Armstrong, Robert and Arnouts, Stéphane and Bahcall, Neta A and Bickerton, Steven and Bosch, James and Bundy, Kevin and Capak, Peter L and Chan, James H H and Chiba, Masashi and Coupon, Jean and Egami, Eiichi and Enoki, Motohiro and Finet, Francois and Fujimori, Hiroki and Fujimoto, Seiji and Furusawa, Hisanori and Furusawa, Junko and Goto, Tomotsugu and Goulding, Andy and Greco, Johnny P and Greene, Jenny E and Gunn, James E and Hamana, Takashi and Harikane, Yuichi and Hashimoto, Yasuhiro and Hattori, Takashi and Hayashi, Masao and Hayashi, Yusuke and Hełminiak, Krzysztof G and Higuchi, Ryo and Hikage, Chiaki and Ho, Paul T P and Hsieh, Bau-Ching and Huang, Kuiyun and Huang, Song and Ikeda, Hiroyuki and Imanishi, Masatoshi and Inoue, Akio K and Iwasawa, Kazushi and Iwata, Ikuru and Jaelani, Anton T and Jian, Hung-Yu and Kamata, Yukiko and Karoji, Hiroshi and Kashikawa, Nobunari and Katayama, Nobuhiko and Kawanomoto, Satoshi and Kayo, Issha and Koda, Jin and Koike, Michitaro and Kojima, Takashi and Komiyama, Yutaka and Konno, Akira and Koshida, Shintaro and Koyama, Yusei and Kusakabe, Haruka and Leauthaud, Alexie and Lee, Chien-Hsiu and Lin, Lihwai and Lin, Yen-Ting and Lupton, Robert H and Mandelbaum, Rachel and Matsuoka, Yoshiki and Medezinski, Elinor and Mineo, Sogo and Miyama, Shoken and Miyatake, Hironao and Miyazaki, Satoshi and Momose, Rieko and More, Anupreeta and More, Surhud and Moritani, Yuki and Moriya, Takashi J and Morokuma, Tomoki and Mukae, Shiro and Murata, Ryoma and Murayama, Hitoshi and Nagao, Tohru and Nakata, Fumiaki and Niida, Mana and Niikura, Hiroko and Nishizawa, Atsushi J and Obuchi, Yoshiyuki and Oguri, Masamune and Oishi, Yukie and Okabe, Nobuhiro and Okamoto, Sakurako and Okura, Yuki and Ono, Yoshiaki and Onodera, Masato and Onoue, Masafusa and Osato, Ken and Ouchi, Masami and Price, Paul A and Pyo, Tae-Soo and Sako, Masao and Sawicki, Marcin and Shibuya, Takatoshi and Shimasaku, Kazuhiro and Shimono, Atsushi and Shirasaki, Masato and Silverman, John D and Simet, Melanie and Speagle, Joshua and Spergel, David N and Strauss, Michael A and Sugahara, Yuma and Sugiyama, Naoshi and Suto, Yasushi and Suyu, Sherry H and Suzuki, Nao and Tait, Philip J and Takada, Masahiro and Takata, Tadafumi and Tamura, Naoyuki and Tanaka, Manobu M and Tanaka, Masaomi and Tanaka, Masayuki and Tanaka, Yoko and Terai, Tsuyoshi and Terashima, Yuichi and Toba, Yoshiki and Tominaga, Nozomu and Toshikawa, Jun and Turner, Edwin L and Uchida, Tomohisa and Uchiyama, Hisakazu and Umetsu, Keiichi and Uraguchi, Fumihiro and Urata, Yuji and Usuda, Tomonori and Utsumi, Yousuke and Wang, Shiang-Yu and Wang, Wei-Hao and Wong, Kenneth C and Yabe, Kiyoto and Yamada, Yoshihiko and Yamanoi, Hitomi and Yasuda, Naoki and Yeh, Sherry and Yonehara, Atsunori and Yuma, Suraphong},
	month = jan,
	year = {2018},
	pages = {S4},
}

@article{ahumada_16th_2020,
	title = {The 16th {Data} {Release} of the {Sloan} {Digital} {Sky} {Surveys}: {First} {Release} from the {APOGEE}-2 {Southern} {Survey} and {Full} {Release} of {eBOSS} {Spectra}},
	volume = {249},
	issn = {0067-0049},
	shorttitle = {The 16th {Data} {Release} of the {Sloan} {Digital} {Sky} {Surveys}},
	url = {https://doi.org/10.3847/1538-4365/ab929e},
	doi = {10.3847/1538-4365/ab929e},
	language = {en},
	number = {1},
	urldate = {2026-07-07},
	journal = {ApJS},
	publisher = {The American Astronomical Society},
	author = {Ahumada, Romina and Prieto, Carlos Allende and Almeida, Andrés and Anders, Friedrich and Anderson, Scott F. and Andrews, Brett H. and Anguiano, Borja and Arcodia, Riccardo and Armengaud, Eric and Aubert, Marie and Avila, Santiago and Avila-Reese, Vladimir and Badenes, Carles and Balland, Christophe and Barger, Kat and Barrera-Ballesteros, Jorge K. and Basu, Sarbani and Bautista, Julian and Beaton, Rachael L. and Beers, Timothy C. and Benavides, B. Izamar T. and Bender, Chad F. and Bernardi, Mariangela and Bershady, Matthew and Beutler, Florian and Bidin, Christian Moni and Bird, Jonathan and Bizyaev, Dmitry and Blanc, Guillermo A. and Blanton, Michael R. and Boquien, Médéric and Borissova, Jura and Bovy, Jo and Brandt, W. N. and Brinkmann, Jonathan and Brownstein, Joel R. and Bundy, Kevin and Bureau, Martin and Burgasser, Adam and Burtin, Etienne and Cano-Díaz, Mariana and Capasso, Raffaella and Cappellari, Michele and Carrera, Ricardo and Chabanier, Solène and Chaplin, William and Chapman, Michael and Cherinka, Brian and Chiappini, Cristina and Doohyun Choi, Peter and Chojnowski, S. Drew and Chung, Haeun and Clerc, Nicolas and Coffey, Damien and Comerford, Julia M. and Comparat, Johan and da Costa, Luiz and Cousinou, Marie-Claude and Covey, Kevin and Crane, Jeffrey D. and Cunha, Katia and Ilha, Gabriele da Silva and Dai, Yu Sophia and Damsted, Sanna B. and Darling, Jeremy and Davidson, James W. and Davies, Roger and Dawson, Kyle and De, Nikhil and de la Macorra, Axel and De Lee, Nathan and Queiroz, Anna Bárbara de Andrade and Deconto Machado, Alice and de la Torre, Sylvain and Dell’Agli, Flavia and du Mas des Bourboux, Hélion and Diamond-Stanic, Aleksandar M. and Dillon, Sean and Donor, John and Drory, Niv and Duckworth, Chris and Dwelly, Tom and Ebelke, Garrett and Eftekharzadeh, Sarah and Davis Eigenbrot, Arthur and Elsworth, Yvonne P. and Eracleous, Mike and Erfanianfar, Ghazaleh and Escoffier, Stephanie and Fan, Xiaohui and Farr, Emily and Fernández-Trincado, José G. and Feuillet, Diane and Finoguenov, Alexis and Fofie, Patricia and Fraser-McKelvie, Amelia and Frinchaboy, Peter M. and Fromenteau, Sebastien and Fu, Hai and Galbany, Lluís and Garcia, Rafael A. and García-Hernández, D. A. and Oehmichen, Luis Alberto Garma and Ge, Junqiang and Maia, Marcio Antonio Geimba and Geisler, Doug and Gelfand, Joseph and Goddy, Julian and Gonzalez-Perez, Violeta and Grabowski, Kathleen and Green, Paul and Grier, Catherine J. and Guo, Hong and Guy, Julien and Harding, Paul and Hasselquist, Sten and Hawken, Adam James and Hayes, Christian R. and Hearty, Fred and Hekker, S. and Hogg, David W. and Holtzman, Jon A. and Horta, Danny and Hou, Jiamin and Hsieh, Bau-Ching and Huber, Daniel and Hunt, Jason A. S. and Chitham, J. Ider and Imig, Julie and Jaber, Mariana and Angel, Camilo Eduardo Jimenez and Johnson, Jennifer A. and Jones, Amy M. and Jönsson, Henrik and Jullo, Eric and Kim, Yerim and Kinemuchi, Karen and Kirkpatrick IV, Charles C. and Kite, George W. and Klaene, Mark and Kneib, Jean-Paul and Kollmeier, Juna A. and Kong, Hui and Kounkel, Marina and Krishnarao, Dhanesh and Lacerna, Ivan and Lan, Ting-Wen and Lane, Richard R. and Law, David R. and Le Goff, Jean-Marc and Leung, Henry W. and Lewis, Hannah and Li, Cheng and Lian, Jianhui and Lin, Lihwai and Long, Dan and Longa-Peña, Penélope and Lundgren, Britt and Lyke, Brad W. and Ted Mackereth, J. and MacLeod, Chelsea L. and Majewski, Steven R. and Manchado, Arturo and Maraston, Claudia and Martini, Paul and Masseron, Thomas and Masters, Karen L. and Mathur, Savita and McDermid, Richard M. and Merloni, Andrea and Merrifield, Michael and Mészáros, Szabolcs and Miglio, Andrea and Minniti, Dante and Minsley, Rebecca and Miyaji, Takamitsu and Mohammad, Faizan Gohar and Mosser, Benoit and Mueller, Eva-Maria and Muna, Demitri and Muñoz-Gutiérrez, Andrea and Myers, Adam D. and Nadathur, Seshadri and Nair, Preethi and Nandra, Kirpal and do Nascimento, Janaina Correa and Nevin, Rebecca Jean and Newman, Jeffrey A. and Nidever, David L. and Nitschelm, Christian and Noterdaeme, Pasquier and O’Connell, Julia E. and Olmstead, Matthew D. and Oravetz, Daniel and Oravetz, Audrey and Osorio, Yeisson and Pace, Zachary J. and Padilla, Nelson and Palanque-Delabrouille, Nathalie and Palicio, Pedro A. and Pan, Hsi-An and Pan, Kaike and Parker, James and Paviot, Romain and Peirani, Sebastien and Ramŕez, Karla Peña and Penny, Samantha and Percival, Will J. and Perez-Fournon, Ismael and Pérez-Ràfols, Ignasi and Petitjean, Patrick and Pieri, Matthew M. and Pinsonneault, Marc and Poovelil, Vijith Jacob and Povick, Joshua Tyler and Prakash, Abhishek and Price-Whelan, Adrian M. and Raddick, M. Jordan and Raichoor, Anand and Ray, Amy and Rembold, Sandro Barboza and Rezaie, Mehdi and Riffel, Rogemar A. and Riffel, Rogério and Rix, Hans-Walter and Robin, Annie C. and Roman-Lopes, A. and Román-Zúñiga, Carlos and Rose, Benjamin and Ross, Ashley J. and Rossi, Graziano and Rowlands, Kate and Rubin, Kate H. R. and Salvato, Mara and Sánchez, Ariel G. and Sánchez-Menguiano, Laura and Sánchez-Gallego, José R. and Sayres, Conor and Schaefer, Adam and Schiavon, Ricardo P. and Schimoia, Jaderson S. and Schlafly, Edward and Schlegel, David and Schneider, Donald P. and Schultheis, Mathias and Schwope, Axel and Seo, Hee-Jong and Serenelli, Aldo and Shafieloo, Arman and Shamsi, Shoaib Jamal and Shao, Zhengyi and Shen, Shiyin and Shetrone, Matthew and Shirley, Raphael and Aguirre, Víctor Silva and Simon, Joshua D. and Skrutskie, M. F. and Slosar, Anže and Smethurst, Rebecca and Sobeck, Jennifer and Sodi, Bernardo Cervantes and Souto, Diogo and Stark, David V. and Stassun, Keivan G. and Steinmetz, Matthias and Stello, Dennis and Stermer, Julianna and Storchi-Bergmann, Thaisa and Streblyanska, Alina and Stringfellow, Guy S. and Stutz, Amelia and Suárez, Genaro and Sun, Jing and Taghizadeh-Popp, Manuchehr and Talbot, Michael S. and Tayar, Jamie and Thakar, Aniruddha R. and Theriault, Riley and Thomas, Daniel and Thomas, Zak C. and Tinker, Jeremy and Tojeiro, Rita and Toledo, Hector Hernandez and Tremonti, Christy A. and Troup, Nicholas W. and Tuttle, Sarah and Unda-Sanzana, Eduardo and Valentini, Marica and Vargas-González, Jaime and Vargas-Magaña, Mariana and Vázquez-Mata, Jose Antonio and Vivek, M. and Wake, David and Wang, Yuting and Weaver, Benjamin Alan and Weijmans, Anne-Marie and Wild, Vivienne and Wilson, John C. and Wilson, Robert F. and Wolthuis, Nathan and Wood-Vasey, W. M. and Yan, Renbin and Yang, Meng and Yèche, Christophe and Zamora, Olga and Zarrouk, Pauline and Zasowski, Gail and Zhang, Kai and Zhao, Cheng and Zhao, Gongbo and Zheng, Zheng and Zheng, Zheng and Zhu, Guangtun and Zou, Hu},
	month = jun,
	year = {2020},
	pages = {3},
}

@misc{collaboration_desi_2016,
	title = {The {DESI} {Experiment} {Part} {I}: {Science},{Targeting}, and {Survey} {Design}},
	shorttitle = {The {DESI} {Experiment} {Part} {I}},
	url = {http://arxiv.org/abs/1611.00036},
	doi = {10.48550/arXiv.1611.00036},
	urldate = {2026-07-07},
	publisher = {arXiv},
	author = {Collaboration, DESI and Aghamousa, Amir and Aguilar, Jessica and Ahlen, Steve and Alam, Shadab and Allen, Lori E. and Prieto, Carlos Allende and Annis, James and Bailey, Stephen and Balland, Christophe and Ballester, Otger and Baltay, Charles and Beaufore, Lucas and Bebek, Chris and Beers, Timothy C. and Bell, Eric F. and Bernal, José Luis and Besuner, Robert and Beutler, Florian and Blake, Chris and Bleuler, Hannes and Blomqvist, Michael and Blum, Robert and Bolton, Adam S. and Briceno, Cesar and Brooks, David and Brownstein, Joel R. and Buckley-Geer, Elizabeth and Burden, Angela and Burtin, Etienne and Busca, Nicolas G. and Cahn, Robert N. and Cai, Yan-Chuan and Cardiel-Sas, Laia and Carlberg, Raymond G. and Carton, Pierre-Henri and Casas, Ricard and Castander, Francisco J. and Cervantes-Cota, Jorge L. and Claybaugh, Todd M. and Close, Madeline and Coker, Carl T. and Cole, Shaun and Comparat, Johan and Cooper, Andrew P. and Cousinou, M.-C. and Crocce, Martin and Cuby, Jean-Gabriel and Cunningham, Daniel P. and Davis, Tamara M. and Dawson, Kyle S. and Macorra, Axel de la and Vicente, Juan De and Delubac, Timothée and Derwent, Mark and Dey, Arjun and Dhungana, Govinda and Ding, Zhejie and Doel, Peter and Duan, Yutong T. and Ealet, Anne and Edelstein, Jerry and Eftekharzadeh, Sarah and Eisenstein, Daniel J. and Elliott, Ann and Escoffier, Stéphanie and Evatt, Matthew and Fagrelius, Parker and Fan, Xiaohui and Fanning, Kevin and Farahi, Arya and Farihi, Jay and Favole, Ginevra and Feng, Yu and Fernandez, Enrique and Findlay, Joseph R. and Finkbeiner, Douglas P. and Fitzpatrick, Michael J. and Flaugher, Brenna and Flender, Samuel and Font-Ribera, Andreu and Forero-Romero, Jaime E. and Fosalba, Pablo and Frenk, Carlos S. and Fumagalli, Michele and Gaensicke, Boris T. and Gallo, Giuseppe and Garcia-Bellido, Juan and Gaztanaga, Enrique and Fusillo, Nicola Pietro Gentile and Gerard, Terry and Gershkovich, Irena and Giannantonio, Tommaso and Gillet, Denis and Gonzalez-de-Rivera, Guillermo and Gonzalez-Perez, Violeta and Gott, Shelby and Graur, Or and Gutierrez, Gaston and Guy, Julien and Habib, Salman and Heetderks, Henry and Heetderks, Ian and Heitmann, Katrin and Hellwing, Wojciech A. and Herrera, David A. and Ho, Shirley and Holland, Stephen and Honscheid, Klaus and Huff, Eric and Hutchinson, Timothy A. and Huterer, Dragan and Hwang, Ho Seong and Laguna, Joseph Maria Illa and Ishikawa, Yuzo and Jacobs, Dianna and Jeffrey, Niall and Jelinsky, Patrick and Jennings, Elise and Jiang, Linhua and Jimenez, Jorge and Johnson, Jennifer and Joyce, Richard and Jullo, Eric and Juneau, Stéphanie and Kama, Sami and Karcher, Armin and Karkar, Sonia and Kehoe, Robert and Kennamer, Noble and Kent, Stephen and Kilbinger, Martin and Kim, Alex G. and Kirkby, David and Kisner, Theodore and Kitanidis, Ellie and Kneib, Jean-Paul and Koposov, Sergey and Kovacs, Eve and Koyama, Kazuya and Kremin, Anthony and Kron, Richard and Kronig, Luzius and Kueter-Young, Andrea and Lacey, Cedric G. and Lafever, Robin and Lahav, Ofer and Lambert, Andrew and Lampton, Michael and Landriau, Martin and Lang, Dustin and Lauer, Tod R. and Goff, Jean-Marc Le and Guillou, Laurent Le and Suu, Auguste Le Van and Lee, Jae Hyeon and Lee, Su-Jeong and Leitner, Daniela and Lesser, Michael and Levi, Michael E. and L'Huillier, Benjamin and Li, Baojiu and Liang, Ming and Lin, Huan and Linder, Eric and Loebman, Sarah R. and Lukić, Zarija and Ma, Jun and MacCrann, Niall and Magneville, Christophe and Makarem, Laleh and Manera, Marc and Manser, Christopher J. and Marshall, Robert and Martini, Paul and Massey, Richard and Matheson, Thomas and McCauley, Jeremy and McDonald, Patrick and McGreer, Ian D. and Meisner, Aaron and Metcalfe, Nigel and Miller, Timothy N. and Miquel, Ramon and Moustakas, John and Myers, Adam and Naik, Milind and Newman, Jeffrey A. and Nichol, Robert C. and Nicola, Andrina and Costa, Luiz Nicolati da and Nie, Jundan and Niz, Gustavo and Norberg, Peder and Nord, Brian and Norman, Dara and Nugent, Peter and O'Brien, Thomas and Oh, Minji and Olsen, Knut A. G. and Padilla, Cristobal and Padmanabhan, Hamsa and Padmanabhan, Nikhil and Palanque-Delabrouille, Nathalie and Palmese, Antonella and Pappalardo, Daniel and Pâris, Isabelle and Park, Changbom and Patej, Anna and Peacock, John A. and Peiris, Hiranya V. and Peng, Xiyan and Percival, Will J. and Perruchot, Sandrine and Pieri, Matthew M. and Pogge, Richard and Pollack, Jennifer E. and Poppett, Claire and Prada, Francisco and Prakash, Abhishek and Probst, Ronald G. and Rabinowitz, David and Raichoor, Anand and Ree, Chang Hee and Refregier, Alexandre and Regal, Xavier and Reid, Beth and Reil, Kevin and Rezaie, Mehdi and Rockosi, Constance M. and Roe, Natalie and Ronayette, Samuel and Roodman, Aaron and Ross, Ashley J. and Ross, Nicholas P. and Rossi, Graziano and Rozo, Eduardo and Ruhlmann-Kleider, Vanina and Rykoff, Eli S. and Sabiu, Cristiano and Samushia, Lado and Sanchez, Eusebio and Sanchez, Javier and Schlegel, David J. and Schneider, Michael and Schubnell, Michael and Secroun, Aurélia and Seljak, Uros and Seo, Hee-Jong and Serrano, Santiago and Shafieloo, Arman and Shan, Huanyuan and Sharples, Ray and Sholl, Michael J. and Shourt, William V. and Silber, Joseph H. and Silva, David R. and Sirk, Martin M. and Slosar, Anze and Smith, Alex and Smoot, George F. and Som, Debopam and Song, Yong-Seon and Sprayberry, David and Staten, Ryan and Stefanik, Andy and Tarle, Gregory and Tie, Suk Sien and Tinker, Jeremy L. and Tojeiro, Rita and Valdes, Francisco and Valenzuela, Octavio and Valluri, Monica and Vargas-Magana, Mariana and Verde, Licia and Walker, Alistair R. and Wang, Jiali and Wang, Yuting and Weaver, Benjamin A. and Weaverdyck, Curtis and Wechsler, Risa H. and Weinberg, David H. and White, Martin and Yang, Qian and Yeche, Christophe and Zhang, Tianmeng and Zhao, Gong-Bo and Zheng, Yi and Zhou, Xu and Zhou, Zhimin and Zhu, Yaling and Zou, Hu and Zu, Ying},
	month = dec,
	year = {2016},
	note = {arXiv:1611.00036 [astro-ph.IM]},
}

@misc{parker_aion-1_2025,
	title = {{AION}-1: {Omnimodal} {Foundation} {Model} for {Astronomical} {Sciences}},
	shorttitle = {{AION}-1},
	url = {http://arxiv.org/abs/2510.17960},
	doi = {10.48550/arXiv.2510.17960},
	urldate = {2026-07-07},
	publisher = {arXiv},
	author = {Parker, Liam and Lanusse, Francois and Shen, Jeff and Liu, Ollie and Hehir, Tom and Sarra, Leopoldo and Meyer, Lucas and Bowles, Micah and Wagner-Carena, Sebastian and Qu, Helen and Golkar, Siavash and Bietti, Alberto and Bourfoune, Hatim and Casserau, Nathan and Cornette, Pierre and Hirashima, Keiya and Krawezik, Geraud and Ohana, Ruben and Lourie, Nicholas and McCabe, Michael and Morel, Rudy and Mukhopadhyay, Payel and Pettee, Mariel and Blancard, Bruno Regaldo-Saint and Cho, Kyunghyun and Cranmer, Miles and Ho, Shirley},
	month = oct,
	year = {2025},
	note = {arXiv:2510.17960 [astro-ph.IM]},
}

@article{lastufka_examining_2025,
	title = {Examining vision foundation models for classification and detection in optical and radio astronomy},
	volume = {703},
	copyright = {© The Authors 2025},
	issn = {0004-6361, 1432-0746},
	url = {https://www.aanda.org/articles/aa/abs/2025/11/aa53691-25/aa53691-25.html},
	doi = {10.1051/0004-6361/202553691},
	language = {en},
	urldate = {2026-07-07},
	journal = {A\&A},
	publisher = {EDP Sciences},
	author = {Lastufka, E. and Bait, O. and Drozdova, M. and Kinakh, V. and Piras, D. and Audard, M. and Dessauges-Zavadsky, M. and Holotyak, T. and Schaerer, D. and Voloshynovskiy, S.},
	month = nov,
	year = {2025},
	pages = {A217},
}

@article{parker_astroclip_2024,
	title = {{AstroCLIP}: {A} {Cross}-{Modal} {Foundation} {Model} for {Galaxies}},
	volume = {531},
	issn = {0035-8711, 1365-2966},
	shorttitle = {{AstroCLIP}},
	url = {http://arxiv.org/abs/2310.03024},
	doi = {10.1093/mnras/stae1450},
	number = {4},
	urldate = {2026-07-10},
	journal = {Monthly Notices of the Royal Astronomical Society},
	author = {Parker, Liam and Lanusse, Francois and Golkar, Siavash and Sarra, Leopoldo and Cranmer, Miles and Bietti, Alberto and Eickenberg, Michael and Krawezik, Geraud and McCabe, Michael and Ohana, Ruben and Pettee, Mariel and Blancard, Bruno Regaldo-Saint and Tesileanu, Tiberiu and Cho, Kyunghyun and Ho, Shirley},
	month = jun,
	year = {2024},
	note = {arXiv:2310.03024 [astro-ph.IM]},
	pages = {4990--5011},
}

@misc{simeoni_dinov3_2025,
	title = {{DINOv3}},
	url = {http://arxiv.org/abs/2508.10104},
	doi = {10.48550/arXiv.2508.10104},
	urldate = {2026-07-13},
	publisher = {arXiv},
	author = {Siméoni, Oriane and Vo, Huy V. and Seitzer, Maximilian and Baldassarre, Federico and Oquab, Maxime and Jose, Cijo and Khalidov, Vasil and Szafraniec, Marc and Yi, Seungeun and Ramamonjisoa, Michaël and Massa, Francisco and Haziza, Daniel and Wehrstedt, Luca and Wang, Jianyuan and Darcet, Timothée and Moutakanni, Théo and Sentana, Leonel and Roberts, Claire and Vedaldi, Andrea and Tolan, Jamie and Brandt, John and Couprie, Camille and Mairal, Julien and Jégou, Hervé and Labatut, Patrick and Bojanowski, Piotr},
	month = aug,
	year = {2025},
	note = {arXiv:2508.10104 [cs.CV]},
}

@misc{leung_henryskygalaxy10_2026,
	title = {henrysky/{Galaxy10}},
	copyright = {MIT},
	url = {https://github.com/henrysky/Galaxy10},
	urldate = {2026-08-29},
	author = {Leung, Henry},
	month = jul,
	year = {2026},
	note = {original-date: 2021-03-16T20:23:19Z},
}

\newpage
\appendix

\section{Appendix}

\subsection{Code}
%Code will be made publicly available upon acceptance. 

Code for unimodal and joint pre-training, as well as evaluation, is available at \url{https://github.com/elastufka/dinospec_repo}.

\subsection{Dataset construction}\label{sec:app_data}

HSC, SDSS, and DESI data were sourced from the Multimodal Universe \citep{} version 1, available at \url{https://users.flatironinstitute.org/~polymathic/data/MultimodalUniverse/v1/}

Because the images in the GZ10 evaluation dataset are $grz$ LegacySurvey images and the AION-1 tokenizer requires $griz$ LegacySurvey images or $grizy$ HSC images as input, we chose to cross-match the GZ10 sources to available HSC data since we already had the required five-channel HSC images on hand. This resulted in a much smaller evaluation dataset than the original GZ10, with only 8 samples of the rarest class. We performed a class-balanced train-test split to ensure that all classes were represented in both the train and test sets. The use of this smaller dataset for evaluation is why the reported F1-score for the AION-1 baseline on the galaxy morphology classification task is lower than what is found in their paper.

Data and labels for the spectral classification and redshift prediction tasks are from \url{https://huggingface.co/datasets/Fucheng/GaSNet-II-SDSS-dataset}

\subsection{Adapter pre-training}

The additional transformer blocks used for multimodal alignment were optionally pre-trained before joint vision--spectra training. The underlying DINOv3 vision backbone and AION-1 spectral tokenizer were kept frozen during this stage. The two vision adapter blocks were initialized with random weights and then pre-trained using the DINOv3 self-distillation objective on the \textit{hsc} dataset. Given student and teacher network outputs $p_s$ and $p_t$, the objective minimizes the cross-entropy between their softened output distributions,

\begin{equation}
\mathcal{L}_{\mathrm{DINO}} =
-\sum_k p_t(k)\log p_s(k),
\end{equation}

where the teacher network is updated using an exponential moving average of the student parameters. After 50 pre-training epochs, the vision model (DINOv3-Large frozen backbone plus two ViT blocks) achieved a F1 score of $0.717 \pm 7.1e-03$ on optical galaxy classification. %The student model is used for multimodal alignment following pre-traning. %This pre-training allows the additional vision blocks to learn domain-specific image representations while maintaining compatibility with the frozen DINOv3 backbone.

The spectral adapter blocks were pre-trained using a masked-token prediction objective analogous to the BERT objective. Spectral tokens produced by the pre-trained AION-1 tokenizer were randomly masked, and the transformer blocks were trained to reconstruct the missing tokens from the surrounding context. The objective is given by

\begin{equation}
\mathcal{L}_{\mathrm{BERT}} =
-\sum_{i\in M}\log P(x_i|\mathbf{x}_{\setminus M}),
\end{equation}

where $M$ denotes the set of masked token positions. Spectral adapter pre-training was performed on the \textit{sdss\_desi} training dataset. A validation dataset was used to prevent overfitting. The 16-block transformer was pre-trained for fewer epochs than the others due to memory constraints. Spectral adapters were pre-trained for longer than the vision adapter because without the the inference cost ($\sim17$ ms/image) of the DINOv3 backbone training proceeded much faster. Table \ref{tab:spectra_pretrain} shows the performance of the pre-trained spectral adapters on the redshift prediction task.

\begin{table}[h]
%\resizebox{\textwidth}{!}{%
\centering
\begin{tabular}{lll}
\toprule
Model            & Pre-training Epochs       & z $R^2$ score  \\ \midrule
%MLP only &  -     &  0.825       \\ %evalz_mlp
%minimal &      &         \\ % evalx_dinov3
2 block Transformer &   500    &    $0.912 \pm 8.1e-03$    \\ % redshift_spec_pretrain_2_512-836324
8 block Transformer &   500    &     $0.897 \pm 9.3e-03$   \\ % evalx_dinov3_52982
16 block Transformer &   100    &     $0.906  \pm 3.4e-03$  \\ % evalx_dinov3_52982
\bottomrule
\end{tabular}
%}
\caption{\label{tab:spectra_pretrain}Spectral adapter pre-training}
\end{table}

The frozen DINOv3-L vision encoder provides a strong baseline for morphology classification, achieving an F1 score of $0.722 \pm 5.2\times10^{-3}$. Adding pre-trained vision adapter blocks without multimodal alignment does not improve performance, indicating that additional visual capacity alone is insufficient. 

\subsection{Additional joint training and evaluation details}

Joint training was performed on paired image--spectrum observations. For each batch of $N$ pairs, the vision and spectral encoders produce L2-normalized representations $\mathbf{v}i$ and $\mathbf{s}i$. Pairwise similarity logits are computed as
$
S{ij} = \exp(\gamma),\mathbf{v}i^\top\mathbf{s}j,
$
where $\gamma$ is a learned logit-scale parameter. We optimize the symmetric cross-entropy CLIP objective,
\begin{equation}
\mathcal{L_{\mathrm{CLIP}}} =
-\frac{1}{2N}
\sum{i=1}^{N}
\left[
\log\frac{e^{S{ii}}}{\sum{j=1}^{N}e^{S_{ij}}}
+
\log\frac{e^{S_{ii}}}{\sum_{j=1}^{N}e^{S_{ji}}}
\right]
\end{equation}

The first term performs image-to-spectrum classification and the second spectrum-to-image classification. Thus, each paired observation forms a positive pair, while the remaining observations in the batch serve as in-batch negatives.

%Images were center-cropped to $96\times96$ pixels and resized to $256\times256$. No additional spatial or color augmentations were used during joint training. The spectral inputs were the pre-tokenized outputs of the AION-1 spectral tokenizer.

Training was performed for 150 epochs using a learning rate of $5\times10^{-6}$, cosine learning-rate decay, a 10\% warmup, weight decay of 0.01, and gradient clipping with a maximum norm of 1. Training used two distributed processes with 16 paired examples per process and four gradient-accumulation steps, corresponding to an effective batch size of 128 paired examples. Depending on the experimental configuration, either the vision or spectral branch was frozen for an initial training period before being jointly optimized; the vision backbone and AION-1 spectral tokenizer remained frozen throughout.

For downstream evaluation, each trained branch was evaluated independently using a 2-layer MLP prediction head. Classification and regression heads were trained separately for each task, with learning rates selected through a small hyperparameter sweep between $5e-6$ and $1e-5$ and early stopping. Table \ref{tab:lrs} shows the learning rates selected for each model and task. Each evaluation was repeated with seven random seeds.

\begin{table*}[h]
\resizebox{\textwidth}{!}{
\begin{tabular}{rlllll}
\toprule
\# & Transformer Blocks & Pre-trained ViT Blocks? & Image lr & Spectra lr & Redshift lr \\
\midrule
1 & 2* & N + LoRA & $5e-5$ & $4e-6$ & $1e-5$ \\
1 & 2* & N & $2.5e-5$ & $5e-6$ & $1e-5$ \\
1 & 16* & N &  $4e-5$ & $2e-6$ & $5e-6$ \\
6 & 16 & N & $2e-5$ & $5e-6$ & $5e-6$ \\
6 & 8 & N & $2e-5$ & $5e-6$ & $1e-5$ \\
1 & 8* & N &  $5e-5$ & $3e-6$ & $7e-6$ \\
%0 & - & N & $0.752 \pm 5.7e-3$ & $0.697 \pm 1.0e-3 $ & $0.910 \pm 4.8e-3$ \\
6 & 2 & N & $4e-5$ & $2e-6$ & $9e-6$ \\
2 & 2 & Y* &  $3e-5$ & $5e-6$ & $6e-6$ \\
4 & 16\^ & Y & $4.5e-5$ & $2e-6$ & $9e-6$ \\
0 & - & Y & $3.5e-5$ & $2e-6$ & $9e-6$ \\
5 & 8 & Y & $2e-5$ & $5e-6$ & $8e-6$ \\
4 & 8\^ & Y & $2e-5$ & $5e-6$ & $9e-6$ \\
5 & 2 & Y & $4e-5$ & $2e-6$ & $8e-6$ \\
3 & 2\^ & Y* & $4e-5$ & $3e-6$ & $1e-5$ \\
2 & 8 & Y* & $3e-5$ & $3e-6$ & $9e-6$ \\
4 & 2\^ & Y & $3.5e-5$ & $4e-6$ & $9e-6$ \\
5 & 16 & Y & $3.5e-5$ & $2e-6$ & $5e-6$ \\
3 & 8\^ & Y* & $4e-5$ & $3e-6$ & $9e-6$ \\
\bottomrule
\end{tabular}
}
\caption{\label{tab:lrs} Learning rates selected for each model and task (galaxy morphology classification, spectral type classification, and redshift prediction) after a small sweep (* frozen model first 50 epochs, \^{}randomly initialized).}
\end{table*}

\subsection{Complete Results}\label{sec:full_results}

In the main text, we only reported results that exceeded the morphology classification baseline of DINOv3-Large in the paper body. Complete results are shown below, including the results using the full Galaxy10 DECaLS dataset using LegacySurvey images. No results on spectral classification or redshift prediction are available for the DINOv3 baselines, because the DINOv3 model applicable to images only.

\begin{table*}[h]
\resizebox{\textwidth}{!}{
\begin{tabular}{rllllll}
\toprule
\# & Transformer  & Pre-trained  & HSC Image F1 & LS Image F1 & Spectra F1 & Spectra R$^2$ \\
& Blocks &  ViT Blocks? &  &  &  &  \\

\midrule
1 & 2* & N + LoRA & $\mathbf{0.796 \pm 5.8e-3}$ & & $0.708 \pm 3.4e-3$ & $0.879 \pm 1.7e-2$ \\
1 & 2* & N & $0.779 \pm 6.1e-3$ & $\mathbf{0.788 \pm 2.7e-3}$ & $0.704 \pm 5.7e-3$ & $0.882 \pm 1.4e-2$ \\
1 & 16* & N & $0.773 \pm 5.5e-3$ & $0.775 \pm 1.9e-3$ & $0.651 \pm 1.4e-3$ & $0.893 \pm 1.0e-3$ \\
6 & 16 & N & $0.768 \pm 5.5e-3$ & $0.773 \pm 1.7e-4$ & $0.659 \pm 3.5e-3$ & $0.891 \pm 5.5e-4$ \\
6 & 8 & N & $0.764 \pm 5.3e-3$ & $0.771 \pm 4.4e-4$ & $0.663 \pm 4.7e-3$ & $0.895 \pm 1.0e-3$ \\
1 & 8* & N & $0.758 \pm 5.3e-3$ & $0.776 \pm 1.3e-3$ & $0.667 \pm 5.5e-3$ & $0.895 \pm 6.2e-4$ \\
%0 & - & N & $0.752 \pm 5.7e-3$ & $0.697 \pm 1.0e-3 $ & $0.910 \pm 4.8e-3$ \\
6 & 2 & N & $0.747 \pm 4.6e-3$ & $0.769 \pm 3.0e-3$  &$0.688 \pm 3.9e-3$ & $0.884 \pm 1.6e-2$ \\
2 & 2 & Y* & $0.732 \pm 4.2e-3$ & $0.772 \pm 1.0e-3$ & $\mathbf{0.736 \pm 2.8e-3}$ & $0.879 \pm 3.9e-3$ \\
4 & 16\^ & Y & $0.731 \pm 6.1e-3$& $0.773 \pm 1.3e-3$  & $0.732 \pm 1.4e-3$ & $0.860 \pm 4.4e-3$ \\
0 & - & Y & $0.731 \pm 6.0e-3$ & $0.773 \pm 7.0e-4$  & $0.697 \pm 1.0e-3 $ & $0.910 \pm 4.8e-3$ \\
5 & 8 & Y & $0.728 \pm 8.6e-3$ & $0.771 \pm 4.4e-4$ & $0.711 \pm 4.5e-3$ & $0.898 \pm 1.1e-3$ \\
4 & 8\^ & Y & $0.728 \pm 6.2e-3$ & $0.771 \pm 2.0e-4$ & $0.717 \pm 2.2e-3$ & $0.891 \pm 5.9e-3$ \\
5 & 2 & Y & $0.724 \pm 5.7e-3$ & $0.772 \pm 1.5e-3$ & $0.706 \pm 2.1e-3$ & $0.890 \pm 3.0e-3$ \\
3 & 2\^ & Y* & $0.724 \pm 5.6e-3$ & $0.768 \pm 2.7e-3$ & $0.714 \pm 5.1e-3$ & $0.873 \pm 2.3e-3$ \\
2 & 8 & Y* & $0.722 \pm 6.0e-3$ & $0.769 \pm 1.5e-3$ & $0.691 \pm 2.1e-2$ & $0.899 \pm 1.2e-3$ \\
4 & 2\^ & Y & $0.721 \pm 4.5e-3$ & $0.772 \pm 1.1e-3$ & $0.709 \pm 4.1e-3$ & $0.888 \pm 2.0e-3$ \\
5 & 16 & Y & $0.720 \pm 5.9e-3$ & $0.772 \pm 7.8e-4$  & $0.653 \pm 3.7e-3$ & $0.884 \pm 1.1e-3$ \\
3 & 8\^ & Y* & $0.717 \pm 8.5e-3$ & $0.764 \pm 5.9e-3$ & $0.731 \pm 2.8e-3$ & $0.874 \pm 6.1e-3$ \\
\midrule
& DINOv3 ViT-L & + LoRA + MLP & $0.757 \pm 1.9e-2$ & - & - & - \\
& DINOv3 ViT-L & + MLP & $0.722 \pm 5.2e-3$ & $0.763 \pm 4.0e-3$ &- & - \\
& AION-1 ViT-B & + MLP & $0.702 \pm 1.2e-2$ & $0.697 \pm 1.0e-3 $ & $0.651 \pm 1.3e-3$  & $\mathbf{0.910 \pm 4.8e-3}$ \\
\bottomrule
\end{tabular}
}
\caption{\label{tab:fullresults}Evaluation results from joint pre-training (* frozen model first 50 epochs, \^{}randomly initialized). Baseline comparisons to vision (DINOv3) and multimodal (AION-1) models included.}
\end{table*}

\end{document}